\documentclass[runningheads]{llncs}
\usepackage{graphicx}
\usepackage{subfig}
\usepackage{hyperref}

\usepackage{tikz,xcolor,hyperref}

\definecolor{lime}{HTML}{A6CE39}
\DeclareRobustCommand{\orcidicon}{%
	\begin{tikzpicture}
	\draw[lime, fill=lime] (0,0) 
	circle [radius=0.16] 
	node[white] {{\fontfamily{qag}\selectfont \tiny ID}};
	\draw[white, fill=white] (-0.0625,0.095) 
	circle [radius=0.007];
	\end{tikzpicture}
	\hspace{-2mm}
}

\foreach \x in {A, ..., Z}{%
	\expandafter\xdef\csname orcid\x\endcsname{\noexpand\href{https://orcid.org/\csname orcidauthor\x\endcsname}{\noexpand\orcidicon}}
}

\begin{document}
\title{The Coronavirus is a Bioweapon: \\ Analysing Coronavirus Fact-Checked Stories}

\titlerunning{The Coronavirus is a Bioweapon}

\author{Lynnette Hui Xian Ng\inst{1}\orcidA \and
Kathleen M. Carley\inst{1}\orcidB}
\authorrunning{Ng, Carley}

\institute{CASOS, Institute for Software Research \\
Carnegie Mellon University, Pittsburgh, PA 15213 \\
\email{\{huixiann, carley\}@andrew.cmu.edu}}

\maketitle              

\begin{abstract}
The 2020 coronavirus pandemic has heightened the need to flag coronavirus-related misinformation, and fact-checking groups have taken to verifying misinformation on the Internet. We explore stories reported by fact-checking groups PolitiFact, Poynter and Snopes from January to June 2020, characterising them into six story clusters before then analyse time-series and story validity trends and the level of agreement across sites. We further break down the story clusters into more granular story types by proposing a unique automated method with a BERT classifier, which can be used to classify diverse story sources, in both fact-checked stories and tweets. 


\end{abstract}

\section{Introduction}
The 2020 coronavirus pandemic has prompted major fact-checking groups like PolitiFact, Poynter and Snopes to verify misinformation on the Internet. Since fact-checking influences citizens' reactions \cite{doi:10.1080/10584609.2014.914613}, coronavirus-related fact checking is crucial in informing the public and reducing panic. Previous works compared election-related misinformation from fact-checking sites. While Amazeen conclude a high level of agreement \cite{doi:10.1080/15377857.2014.959691}, Lim augmented the result \cite{lim2018checking}, finding rare agreement on ambiguous statements. Hossain et. al trained word representations of manually classified COVID-19 tweets before performing automatic classification using ROBERTa and BERTScore \cite{hossainetal}.

This paper classifies coronavirus-related fact-checks into story clusters and empirically analyse their characteristics. We invited human annotators to classify one-third of the stories into more granular story types. We developed an automated method to characterising story types and extended the pipeline to classifying tweet story types, suggesting a semi-supervised way of identifying story types in diver

\section{Data and Methododology}

\subsection{Data Collection}
We collected 6731 fact-checked stories from three well-known main fact checking websites: Poynter\footnote{\url{https://www.poynter.org/ifcn-covid-19-misinformation/}}, Snopes\footnote{\url{https://www.snopes.com/fact-check/}} and PolitiFact\footnote{\url{https://www.politifact.com}} in the timeframe of January 14 2020 to June 5 2020. The stories collected are in the English language. Poynter is part of the International Fact Checking Network, and hosts a coronavirus fact-checking section with over 7000 stories specific to the pandemic. As such, we collected our stories from Poynter from its coronavirus-specific section. PolitiFact is a US-based independent fact checking agency that has a primary focus on politician claims. Snopes is an independent publication that is focused on urban legends, hoaxes and folklores. Tables \ref{tab:factcheckingsummary} and \ref{tab:datafields} describe the dataset.

\begin{table}
\vspace{-4mm}
\centering
\caption{\label{tab:factcheckingsummary}Summary of Stories}
\begin{tabular}{ |p{5cm}|p{5cm}| } 
\hline
 \textbf{Fact Checking Site} & \textbf{Number of Stories} \\ \hline
 Poynter \newline (coronavirus misinformation) & 6139 \\ \hline
 Snopes & 151 \\ \hline
 PolitiFact & 441 \\ \hline
\end{tabular}
\vspace{-6mm}
\end{table}

\begin{table}
\vspace{-6mm}
\centering
\caption{\label{tab:datafields}Data Fields}
\begin{tabular}{ |p{3cm}|p{7cm}| } 
 \hline
 \textbf{Data Field} & \textbf{Explanation} \\ \hline
 Article Id & Unique ID, if given by the website; otherwise self-generated \\ \hline
 Date Reported & Date of story if available; otherwise date the story was highlighted \\ \hline 
 Validity & Truthfulness of the story  \\ \hline 
 Story & Story to be fact checked \\ \hline
 Elaboration & Elaboration to the validity of the story \\ \hline
 Medium & Medium where the story was originated (i.e. Facebook, Twitter, WhatsApp) \\ \hline
\end{tabular}
\vspace{-6mm}
\end{table}

\subsection{Data Preprocessing} 
\textbf{Harmonising Originating Medium.} Each story is tagged with an originating medium where the post was first submitted to the fact-checking site. We first identified top-level domains like \url{.net, .com} and labelled the originators of these claims as ``Website". For the other stories, we perform entity extraction using StanfordNLP Named-Entity Recognition package \cite{10.3115/1219840.1219885} on the originating field and labelled positive results as ``Person". Finally, we parsed the social media platforms that are listed in the originating field and tagged the story accordingly. We harmonise the originating mediums across the sites. A story may have multiple originators, i.e. a story may appear on both Twitter and Facebook.

\textbf{Harmonising Validity.} Given that each website expresses the validity of the stories in different manners, we performed pre-processing on the stories' validity to summarise the categories into: True, Partially True, Partially False, False and Unknown. Table \ref{tab:harmonisingvalidity} shows the harmonisation metric used.

\begin{table}
\vspace{-4mm}
\centering
\caption{\label{tab:harmonisingvalidity}Harmonisation Metric for Story Validity}
\begin{tabular}{ |p{2.3cm}|p{5cm}|p{5cm}| } 
 \hline
\textbf{Harmonised \newline Validity} & \textbf{Explanation} & \textbf{Variations on Fact-Checking Sites} \\ \hline
True & Can be verified by trusted source \newline (eg CDC, peer-reviewed papers) & Correct, Correct Attribution, True \\ \hline
Partially True & Contains verifiable true facts \newline and facts that cannot be verified & Half true, Half truth, \newline Mixed, Mixture, Mostly True, Partially True, Partly True, Partially correct, True but \\ \hline
Partially False & Contains verifiable false facts \newline and facts that cannot be verified & Mostly False, \newline Partly False, Partially False, Two Pinocchios \\ \hline
False & Can be disputed or has been disputed false by trusted source \newline or the organisation/ person in the claim & False, Falseo, Fake, Misleading, \newline Pants on fire, Pants-fire, Scam, Barely-true \\ \hline
Unknown & Cannot be verified or disputed & Org. doesn't apply rating, In dispute, \newline No evidence, Unproven, Unverified, Suspicions \\ \hline
\end{tabular}
\vspace{-4mm}
\end{table}

\textbf{Word Representations.} \label{sec:wordrep} We first perform text preprocessing functions on the story text such as special characters removal, stemming and lemmatization, then construct contextual word embeddings of each story in two different ways: (1) a Bag-Of-Words (BOW) static vector representation using word tokens from Sklearn Python package\cite{sklearn_api}, and (2) a BERT vector representation for contextualised word embeddings using the pre-trained uncased English embedding model from HuggingFace SentenceTransformer \cite{reimers-2020-multilingual-sentence-bert}. 

The BOW vector representation first creates a vector for each sentence that represents the count of word occurrences in each sentence. It can be enhanced by a weighting scheme called TF-IDF to reflect how important the word is to the corpus of sentences.  The BERT representation builds a language transformer model based on the concept that similar words have similar contexts, reflected in that these vectors are closer to each other. 

\subsection{Cluster Analysis on Stories}
Automatic clustering of stories is used to discover a hidden grouping of story clusters. We reduce the dimensions of the constructed story embeddings using Principal Component Analysis before performing kmeans clustering to obtain an automatic grouping of stories. For the rest of our analysis, we segment the stories into these clusters, providing an understanding of each of the story cluster.

\textbf{Classification of Story Validity.} For each cluster, we divided the stories into an 80-20 train-test ratio to construct a series of machine learning models predicting the validity of the story. We represented the story clusters in the two different ways elaborated in the Word Representation paragraph (Paragraph \ref{sec:wordrep}) and compared the classification performances using naive bayes and logsitic regression classifiers. 

\textbf{Level of Agreement across Fact-Checking Sites.} A single story may be reported on multiple sites with slightly different validity. For each cluster, we look at stories across the sites by comparing their BERT embeddings through cosine distance. We find the five closest embeddings above a threshold of 70\%, and take the mode of the reported story validity. If the story validity is a match, we consider the story to have been agreed between both sites.

\subsection{Story Type Categorisation.} Automatic clustering of stories reveal that several story types may be grouped together into a single cluster, and several clusters may contain the same story type. As such, we also categorised stories via manual annotations. We enlisted three annotators between the ages of 25 and 30, who have had exposure to online misinformation on the coronavirus and speak English as their first language. Inter-annotator agreement is resolved by taking the mode of the annotations. These annotators categorised 2000, or one-third, of stories into several categories: Case Occurrences, Commercial Activity/ Promotion, Conspiracy, Correction/ Calling Out, Emergency Responses, Fake Cures, Fake/True Fact or Prevention, Fake/True Public Health Responses and Public Figures. 

We tested two categorisation techniques: (1) Bag-Of-Words (BOW) classifier and (2) BERT classifier. The BOW classifier is constructed from word token representations of the sentences. Instead of enhancing the representation with TF-IDF as in the Classification of Story Validity, we picked out salient entities in each story category. We extract persons from the story using StanfordNLP Named-Entity Recognition package \cite{10.3115/1219840.1219885}. Using extracted person names, we query Wikipedia using the MediaWiki API, and classify the story as a ``Political/ Public Figure" if the person has a dedicated page. For stories without political/ public figures, we check if they contain a predefined list of words relating to each story type. For example, the ``Conspiracy" story type typically contains words like ``bioweapon" or ``5G". The BERT classifier is constructed by matching the story embedding with the embeddings of manually annotated stories. The story type is annotated with the story type of the closest embedding found through smallest cosine distance.


We extend this process to classify 4573 hand-annotated tweets that contain misinformation obtained from manual filtering and annotation of tweets collected with the hashtag \#covid19 using the Twitter Streaming API, and perform cross-comparison against the stories.

\section{Results and Discussion}
Our findings characterise story clusters in fact-checking sites surrounding the 2020 coronavirus pandemic. In the succeeding sections, we present an analysis of the story clusters in terms of the validity of facts, storyline duration and describe level of agreement between fact checking sites. We also present comparisons between automated grouping of stories and manual annotations.

\subsection{Story Clusters}
Each story is represented as a word vector using BERT embeddings, and further reduced to 100 principal components using Principal Component Analysis, capturing 95\% of the variance. Six topics were chosen for kmeans clustering based on the elbow rule from the values of Within-Cluster-Sum of Squared Errors (WSS). The clusters are then manually interpreted. Every story was assigned to a cluster number based on their Euclidean distance to the cluster center in the projected space. We note that some clusters remain internally mixed and most clusters contain multiple story types, and will address the problem in the Section \ref{sec:storyclassification}. 

The story clusters generated from clustering BERT story embeddings mimic human curated storylines from CMU's CASOS Coronavirus website \cite{casos}. The human curated storylines are referenced for manual interpretations of the story clusters. In addition, story clusters also mimic the six misinformation categories manually curated by the CoronavirusFactsAlliance, pointing that misinformation around coronavirus revolve around the discovered story clusters \cite{nature}. Stories are evenly distributed across the story clusters.

\textbf{Story Cluster 1: Photos/Videos, Calling Out/ Correction.} Accounting for about 23\% of the stories, this first topic generally describes stories that contains photos and videos, and stories answering questions about the coronavirus. This topic has been active since January 30, when the pandemic hit and Poynter formed the coronavirus fact checking alliance. Sample stories include: ``Video of man eating bat soup in restaurant in China", and ``Scientists and experts answer questions and rumors about the coronavirus". 

\textbf{Story Cluster 2: Public Figures, Conspiracy/ Prediction.} Accounting for around 20\% of the stories, the second topic was active as early as January 29, which mentioned public figures like celebrities and politicians, conspiracy theories about the source of the coronavirus and past predictions about a global pandemic. Sample stories include: ``Did Kim Jong Un Order North Korea First Coronavirus Patient To Be Executed", ``Did Nostradamus Predict the COVID-19 Pandemic", ``Studies show the coronavirus was engineered to be a bioweapon".


\textbf{Story Cluster 3: False Public Health Responses, Natural Cures/ Prevention.} 12\% of stories fell into the third topic, beginning from January 31, but began to dwindle by April. Sample stories include: ``The Canadian Department of Health issued an emergency notification recommending that people keep their throats moist to protect form the coronavirus", ``Grape vinegar is the antidote to the coronavirus", ``Vitamin C with zinc can prevent and treat the infection".


\textbf{Story Cluster 4: Social Incidents, Commercial Activity/ Promotion, Emergency Responses, False Public Health Responses.} The fourth topic accounts for 12\% of the stories, beginning on January 29 and ending on April 6. Sample stories include: ``Kuwaitt boycotted the products of the Saudi Almari Company", ``20 million Chinese convert to Islam, and the coronavirus does not affect Muslims", ``No, Red Cross is not Offering Coronavirus Home Tests", ``If you are refused service at a store for now wearing a mask call the department of health and report the store".

\textbf{Story Cluster 5: Fake Cures/Vaccines, Fake Facts.} 17\% of the stories fall into the fifth topic, from March 16 to April 9, discussing cures and vaccines and other false facts about the coronavirus. Sample stories include: ``There is magically already a vaccine available", ``COVID-19 comes from rhino horns."

\textbf{Story Cluster 6: Public Health Responses.} 16\% of the stories fall into the final topic, which contains stories on public health responses from February 3 to May 14. Sample stories include: ``Google has donated 59 billion (5900 crores) rupees to fight coronavirus to India", ``China built a hospital for 1,000 people in 10 days and everyone cheered".

\begin{figure}[!tbp]
\vspace{-6mm}
  \centering
  \subfloat[Story Clusters per Website]{\includegraphics[width=0.5\textwidth]{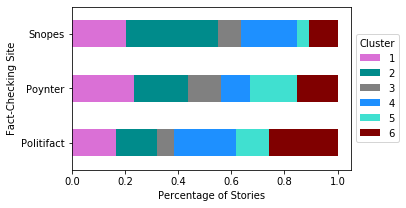}\label{fig:storyclusterspersite}}
  \hfill
  \subfloat[Story Clusters per Originating Medium]{\includegraphics[width=0.5\textwidth]{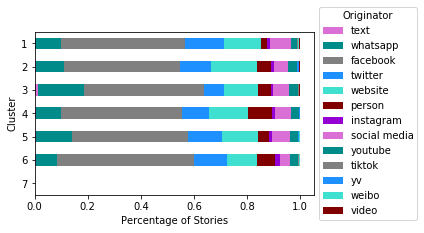}\label{fig:storyclusterperoriginator}}
  \hfill
  \subfloat[Story Clusters Across Time]{\includegraphics[width=0.5\textwidth]{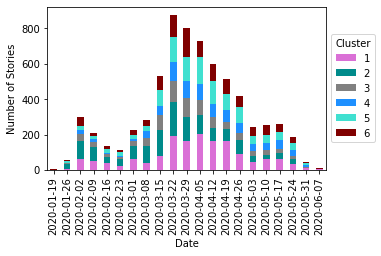}\label{fig:storyclusteracrosstime}}
  \hfill
  \subfloat[Story Clusters Across Validity]{\includegraphics[width=0.5\textwidth]{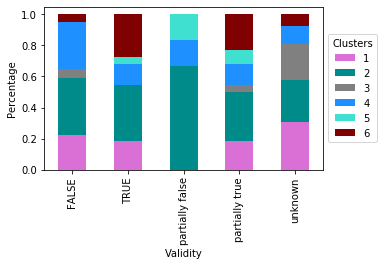}\label{fig:storyclusteracrossvalidity}}
  \label{fig:storyclusters}
  \caption{Story Clusters}
\vspace{-6mm}
\end{figure}

In Figure \ref{fig:storyclusterspersite}, we observe that Snopes has a large proportion of stories in clusters 1 and 2. This is consistent with Snopes' statement on checking folklore and hoaxes, most of which are presented in photos, videos, conspiracy theories and prediction stories. PolitiFact heavily fact checks on cluster 6, looking into claims relating to public health responses made by governments, consistent with their mission to fact-check political claims. The distribution of stories across Poynter is fairly even, likely due to their large network of fact-checkers across many countries. Facebook and WhatsApp are the greatest originating medium of stories across all story clusters (Figure \ref{fig:storyclusterperoriginator}). True stories generally involve public health responses (Figure \ref{fig:storyclusteracrossvalidity}), while partially true stories have a large proportion mentioning public figures.

From the time series chart in Figure \ref{fig:storyclusteracrosstime}, the number of stories increased steadily across the months of February and peaked in end-March. In March, the World Health Organisation declared a global pandemic, many cities and states issued lockdown orders. These global events might have contributed to the sharp increase in stories. 


\subsection{Classification of Story Validity}
\label{sec:storyvalidity}
In classifying story validity, we enhanced the BOW representation with the TF-IDF metric and trained classifiers with Naive Bayes, Support Vector Machines (SVM) and Logistic Regression. We compared this classification technique against constructing BERT vector embeddings on the stories and classifying them using SVM and Logistic Regression.
Table \ref{tab:validity} details the F1 Scores of story validity. There is no significant difference in accuracy whether using a bag-of-words model or a vector-based model, with a good accuracy of 87\% on average. In general, stories in clusters 1 (photos/videos, calling out/correction) and 5 (fake cures/vaccines, fake facts) perform better in the classification models, which could be attributed the presence of unique words, i.e. stories on fake cures tend to contain the words ``cure" and ``vaccines". Stories in clusters 3 (false public health responses, natural cures/ prevention) and 4 (social incidents, commercial activity, false public health responses) performed the worst, because these clusters contain a variety of stories with differing validity.

\begin{table}
\vspace{-7mm}
\centering
\caption{\label{tab:validity}F1 Scores of Story Validity}
\begin{tabular}{ |p{1.5cm}|p{2cm}|p{2cm}|p{2cm}|p{2cm}|p{2cm}| } 
 \hline
\textbf{Cluster} & \textbf{BOW + \newline Naive Bayes} & \textbf{BOW + \newline SVM} & \textbf{BOW + \newline Logistic Regression} & \textbf{BERT + \newline SVM} & \textbf{BERT + \newline Logistic Regression} \\ \hline
1 & 0.90 & 0.90 & 0.92 & 0.92 & 0.90 \\ \hline
2 & 0.85 & 0.86 & 0.88 & 0.85 & 0.85 \\ \hline
3 & 0.82 & 0.83 & 0.86 & 0.82 & 0.84 \\ \hline
4 & 0.85 & 0.88 & 0.88 & 0.85 & 0.84 \\ \hline
5 & 0.90 & 0.90 & 0.90 & 0.90 & 0.89 \\ \hline
6 & 0.87 & 0.87 & 0.88 & 0.85 & 0.88 \\ \hline
Avg & 0.87 & 0.87 & 0.89 & 0.87 & 0.87 \\ \hline
\end{tabular}
\vspace{-7mm}
\end{table}

\subsection{Level of Agreement across Fact Checking Sites}
The level of agreement across the three sites are cross tabulated in Table \ref{tab:agreement}. In particular, we note that the story matches for Story Clusters 4 and 5 are close to 0, and that PolitiFact and Poynter have the highest level of agreement of their stories averaging a 78\% agreement across their stories. Both sites are closely related since Poynter acquired PolitiFact in 2018 \cite{politifact} and thus have overlapping resources, leading to larger proportion of similar stories and agreement.

\begin{table}
\vspace{-7mm}
\centering
\caption{\label{tab:agreement}Level of Agreement Across Fact Checking Sites}
\begin{tabular}{ |p{2cm}|p{3.4cm}|p{3.4cm}|p{3.4cm}| } 
 \hline
\textbf{Cluster} & \textbf{Snopes x PolitiFact} & \textbf{Snopes x Poynter} & \textbf{PolitiFact x Poynter} \\ \hline
1 & 0.04 & 0.26 & 0.70 \\ \hline
2 & 0.22 & 0.31 & 0.47 \\ \hline
3 & 0.13 & 0.17 & 0.70 \\ \hline
4 & 0.10 & 0.00 & 1.00 \\ \hline
5 & 0.00 & 0.00 & 1.00 \\ \hline
6 & 0.02 & 0.04 & 0.94 \\ \hline
Avg & 0.085 & 0.13 & 0.80\\ \hline
\end{tabular}
\vspace{-7mm}
\end{table}

\subsection{Story Type Categorisation}
\label{sec:storyclassification}
We further classify the story clusters into more granular story types, and extended the pipeline to match tweets with misinformation. One-third of the dataset was manually annotated as a ground truth for comparison. Due to the different nature of the misinformation in stories and tweets, human annotators have determined 14 classification types for stories and 16 types for tweets. 

In the Bag-Of-Words classifier, we extracted person names from the stories, and determined if they are public figures by the presence of a Wikipedia page. 19\% of the stories and 14\% of tweets contained a public figure with a dedicated Wikipedia page. In comparing BOW against BERT classifiers, we find that BERT classifiers outperform BOW classifiers, indicating that contextualising word vectors perform better than identifying individual words which may be used in a variety of contexts in the stories. The full results are presented in Table \ref{tab:storytypeaccuracy}.

With the BERT classifier, the classes the best perform are: case occurrences and public figures for stories trained on stories; conspiracy and fake cure for stories trained on tweets; conspiracy and public figures for tweets trained on stories; and conspiracy and panic buying for tweets trained on tweets. From these results, it implies that augmenting the stories with additional information such as presence of a dedicated Wikipedia page does not improve accuracy. We also note that the classifier performs best when classifying the same medium of story types, i.e. stories trained on stories and tweets trained on tweets. In fact, the classification framework performs worse than the random baseline when trained on a different medium of data. This is likely due to the differences in the text structures of each medium. 

From our experiments, we demonstrate the novelty of using the same algorithm based on BERT embeddings that can be used to categorise stories in diverse media. In our experiments, we performed training by manually annotating 33\% of the story types, then perform classification on the same medium type. In all variations of story/tweet categorisation, when trained on the same medium of data (i.e. classifying stories with embeddings trained on stories and tweets with embeddings trained on tweets), our framework correctly classified an average of 59\% and 43\% stories and tweets respectively, which is 4.5 and 2.7 times more accurate than random baseline. Classifying tweets based on story embeddings performed the worst overall because there are story types annotated in tweets that do not appear in stories. These results demonstrate that story type classification is a difficult task and this accuracy is an acceptable improvement over the random baseline. 

\begin{table}
\vspace{-6mm}
\centering
\caption{\label{tab:storytypeaccuracy}Accuracy Scores of Story Type Classification}
\begin{tabular}{ |p{6.5cm}|p{2cm}|p{1.7cm}|p{1.7cm}|} 
 \hline
& \textbf{Precision} & \textbf{Recall} & \textbf{F1-Score} \\ \hline 
\textbf{Stories trained on Stories (BERT)} & 0.59 & 0.59 & 0.58 \\ \hline
\textbf{Stories trained on Stories (BOW)} & 0.56 & 0.43 & 0.48 \\ \hline
\textbf{Stories trained on Tweets (BERT)} & 0.10 & 0.11 & 0.09 \\ \hline
\textbf{Stories trained on Tweets (BOW)} & 0.06 & 0.05 & 0.05 \\ \hline
\textbf{Tweets trained on Stories (BERT)} & 0.12 & 0.14 & 0.13 \\ \hline
\textbf{Tweets trained on Stories (BOW)} & 0.07 & 0.03 & 0.05 \\ \hline
\textbf{Tweets trained on Tweets (BERT)} & 0.43 & 0.43 & 0.43 \\ \hline 
\textbf{Tweets trained on Tweets (BOW)} & 0.35 & 0.22 & 0.27 \\ \hline
\textbf{Stories Random Baseline} & 0.12 & 0.12 & 0.12 \\ \hline
\textbf{Tweets Random Baseline} & 0.16 & 0.16 & 0.16 \\ \hline
\end{tabular}
\vspace{-6mm}
\end{table}

\subsection{Limitations and Future Work}
Several challenges were encountered in the analysis we conducted. The dataset necessitated painstaking pre-processing procedures for textual analysis as each fact-checking site had its own rating scale for story validity. Within the same site, because the posts are written by a variety of authors, each author has his own creative way of expressing story validity. For example, Poynter authors may denote a false claim as ``Pants on fire" or ``Two Pinocchios". The data has an overwhelming percentage of False facts, which results in high recall rates for the classifiers constructed in Section \ref{sec:storyvalidity}. 

Human annotators classify story types based on their inherent knowledge of the situation. In this work, we have enhanced the story information through searching Wikipedia for extracted person's names and predefined lists of words for each story type for our BOW classifier. With contextualised vector representations with BERT outperforming BOW classifiers, promising directions involve further enhancing the story information through verified information.

\section{Conclusion} 
In this paper, we examined coronavirus-related fact-checked stories from three well-known fact-checking websites, and automatically characterised the stories into six clusters. We obtain an average accuracy of 87\% in supervised classification of story validity. By comparing BERT embeddings of the stories across sites, PoltiFact and Poynter has the highest amount of similarity in stories. We further characterised story clusters into more granular story types determined by human annotators, and extended the classification technique to match tweets with misinformation, demonstrating an approach where the same algorithm can be used for classifying different media. Story type classification results perform best when trained on the same medium, of which at least one-third of the data were manually annotated. Contextualised BERT vector representations outperforms a classifier that augments stories with additional information. Our framework correctly classified an average of 59\% and 43\% stories and tweets respectively, which is 4.5 and 2.7 times more accurate than random baseline. 

\bibliographystyle{splncs04}
\bibliography{biblography}
\end{document}